# Responses of Pre-transitional Materials with Stress-Generating Defects to External Stimuli: Superelasticity, Supermagnetostriction, Invar and Elinvar Effects

Wei-Feng Rao[1]*, Ye-Chuan Xu[1], John W. Morris Jr.[2], Armen G. Khachaturyan[2,3]

[1]Department of Materials Physics, and IEMM, Nanjing University of Information Science and Technology, Nanjing 210044, China; [2]Department of Materials Science and Engineering, University of California, Berkeley, CA, 94720, USA; [3]Department of Materials Science and Engineering, Rutgers University, Piscataway, NJ, 08854, USA;

## Abstract:

We considered a generic case of pre-transitional materials with static stress-generating defects, dislocations and coherent nano-precipitates, at temperatures close but above the starting temperature of martensitic transformation, $M_s$. Using the Phase Field Microelasticity theory and 3D simulation, we demonstrated that the local stress generated by these defects produces equilibrium nano-size martensitic embryos (MEs) in pre-transitional state, these embryos being orientation variants of martensite. This is a new type of equilibrium: the thermoelastic equilibrium between the MEs and parent phase in which the total volume of MEs and their size are equilibrium internal thermodynamic parameters. This thermoelastic equilibrium exists only in presence of the stress-generating defects. Cooling the pre-transitional state towards $M_s$ or applying the external stimuli, stress or magnetic field, results in a shift of the thermoelastic equilibrium provided by a reversible anhysteretic growth of MEs that results in a giant ME-generated macroscopic strain. In particular, this effect can be associated with the diffuse phase transformations observed in some ferroelectrics above the Curie point. It is shown that the ME-generated strain is giant and describes a superelasticity if the applied field is stress. It describes a super magnetostriction if the martensite (or austenite) are ferromagnetic and the applied field is a magnetic field. In general, the material with defects can be a multiferroic with a giant multiferroic response if the parent and martensitic phase have different ferroic properties. Finally the ME-generated strain may explain or, at least, contribute to the Invar and Elinvar effects that are typically observed in pre-transitional austenite. The thermoelastic equilibrium and all these effects exist only if the interaction between the defects and MEs is infinite-range.

# Introduction

Functional materials with large responses to externally applied fields, such as stress, electric or magnetic fields, are of great interests for sensors, actuators and many other applications.[1,2] Homogeneous crystals usually have comparatively low intrinsic responses to applied field unless this the material undergoes the field-induced phase transformation. However, the response in this case is usually largely hysteretic or even non-recoverable. For example, the magnetic field-induced fcc→bcc martensitic transformation (MT) in Fe-Ni alloys [3] can generate a transformation strain of the order of Bain strain (~10%) which would be unprecedented magnetostriction of ~100 times larger than that of the champion materials, Terfenol-D. [4] But this transformation is highly hysteretic, not recoverable, and requires prohibitively high magnetic field. A physical reason for these shortcomings is the very high energy cost of martensitic nucleation caused by large lattice misfit between the austenite and martensite, which is also the very reason for producing giant response.

The parent phase above $M_s$, which for brevity we call austenite for any system, is often not homogeneous. There are numerous experimental observations reporting nanoscale tweed-like patterns in transmission electron microscopy (TEM),[5-7] with anomalous thermal, acoustic, elastic properties.[8-11] A particular example is the elastic softening of ferromagnetic shape memory alloy (FSMA) in pre-transitional state, Fe-30at%Pd,[6,12] where the shear modulus undergoes a 7-fold reduction. The tweed-like distribution of structural inhomogeneities observed in this alloy suggests that these inhomogeneities may play an important role in determining its macroscopic responses to external fields. These observations give a hope for achieving giant low-hysteretic or anhysteretic responses in the pre-transitional austenite with nano-scale structural heterogeneities.

The formation of distributions of nano-domains of the martensitic orientation variants forming a tweed-like structure in pre-transitional austenite is a well-known phenomenon. It has been called as tweed structure,[5-7,13] strain glass,[14,15] or distribution of nanodisturbances[16] etc. However, the origin of this complex microstructure and its properties is still controversial.

There are several computer simulations of the pre-transitional state that were performed to resolve this problem.[17-21] They considered a 2D model of a displacive transformation in a pre-transitional state with randomly distributed point defects. The model was based on the simplest approximation in which the point defects were interpreted as spikes of the temperature of the displacive phase transformation. No strain generated by the defects was considered. In spite of all these simplifications, these computer simulations did demonstrate that a presence of such point defects may produce heterogeneous nano-structured domains of the martensitic phase. It was also demonstrated that the entire system of nano-domains can be regarded as a special phase.[19-21] The same 2D model of point defects was used to investigate the strain response of the paraelectric system in 2D mixed state to the applied electric field at temperatures above the ferroelectric Curie temperature.[19-21] It was found that the strain (piezoelectric) response of the ferroelectric/paraelectric mixed state is fully recoverable but, unfortunately, highly hysteretic.

In our study, we shifted a focus of our research considering the thermodynamics and the thermodynamic properties of the 3D pre-transitional state with two types of the most common stress-generating static defects, dislocations and coherent nano-precipitates. We presumed that the stress-generating defects should produce a local stress-induced MT causing the formation of nano-embryos of orientation variants of martensite distributed within the pre-transitional austenite matrix with otherwise stable austenite crystal lattice. We used the Phase Field Microelasticity (PFM) modeling to study the morphology of a mixed state formed by these embryos as well as its thermodynamic, mechanical, thermal, and magnetomechanic properties.

Following Olson and Cohen,[22] we classified these defects in terms of their potency. The defects of weak potency are the ones that were introduced by Olson and Cohen[22] to explain the fluctuation-assisted nucleation of the martensite in the metastable austenite *below* the thermodynamic equilibrium temperature of two phases, $T_0$. However, in this study we considered stress-generating defects of higher potency that could explain the formation of the equilibrium nano-size MEs not only below $T_0$ but also *above* the $M_s$ (in the pre-transitional austenite). Such defects can also explain a vanishing of the nucleation barrier for the formation of the MEs, this is, the effect required for the anhysteretic response to the applied field.

The potency of defects depends not only on the magnitude of the stress generated by the defects and their density, but also on the temperature: the closer the system to $M_s$, the higher the driving force for the MEs formation. For the cases of high potency defects, nano-sized MEs can be produced even without thermal fluctuations. These MEs are equilibrium particles bounded to the defects and unable to exist independently. Their sizes and their total volume should assume equilibrium values at given temperature and applied field. Application of external field in such cases should shift the thermoelastic equilibrium increasing or reducing the total volume fraction of MEs. Lifting of the field should result in a recovery of the initial state.

According to the elasticity theory,[23,24] the formation of coherent MEs should generate the strain consisting of the heterogeneous and homogeneous parts, the homogeneous strain being linear proportional to the volume fraction of the MEs, $\omega_{ME}$, i.e., the macroscopic strain induced by the volume change of MEs under applied fields could be evaluated as $\Delta\bar{\varepsilon}^{ME} \sim \varepsilon_0 \Delta\omega_{ME}$ where $\varepsilon_0$ is a typical misfit strain between the martensite and austenite lattices and $\Delta\omega_{ME}$ is the change of the volume fraction $\omega_{ME}$ caused by the applied field. Therefore, changes of the volume fraction of the MEs caused by the applied field should also change the homogeneous elastic strain, $\Delta\bar{\varepsilon}$. Because the field-induced $\Delta\omega_{ME}$ can reach a significant value and $\varepsilon_0$ of a MT is usually very large (~10%), the strain response, $\Delta\bar{\varepsilon}^{ME} \sim \varepsilon_0 \Delta\omega_{ME}$, could be giant in comparison to that of a homogeneous austenite. Therefore, the described *defect-induced mixed state* formed by MEs is in a special *nano-embryonic* thermoelastic equilibrium in which the distribution of MEs behaves as an aggregate with collective properties. Such a pre-transitional defected austenite could have giant and anhysteretic responses to applied fields leading to (i) superelasticity if the applied field is stress, (ii) supermagnetostriction if the applied field is magnetic and one or both phases are ferromagnetic, and (iii) giant piezoelectricity if the applied field is electric and one or both phases are ferroelectric. It should be emphasized that the response should not necessarily be a strain. *It could be even multiferroic if the austenite and martensite have different ferroic properties.*

In principle, the thermoelastic equilibrium discovered by Kurdyumov and Hundros about 80 years ago[25] is a well-known phenomenon of equilibrium coexistence of two macroscopic volumes of elastically coherent martensite and austenite below the $M_s$ temperature. This special type of equilibrium provides a gradually increase of the volume fraction of the martensitic phases upon cooling caused by a shift of this equilibrium in favor of the martensite. Our paper is the first study discovering the thermoelastic equilibrium that, unlike the conventional one, exists above the $M_s$ temperature (in the stability field of the austenite) and in which the equilibrium martensitic phase exists in the form of nano-size particle rather than macroscopic volumes.

In this paper, we used the realistic 3D PFM theory and computational modeling of the thermodynamics and behavior of the pre-transitional defected austenite, which explicitly takes into account the strain energy contribution to the transformation thermodynamics. We discovered the new effect, an existence of the thermoelastic equilibrium in the pre-transitional austenite that allowed us to predict a possibility of the superelasticity and supermagnetostriction caused by shifting this equilibrium by applied external stress or magnetic fields, respectively. We also discussed how the shift of the same nano-embryonic equilibrium upon altering temperature may causes the Invar and Elinvar effects. In particular, we studied the role of potency of two types of the most typical defects, dislocations and coherent precipitates, as well as the role of the closeness of austenite to the $M_s$. The most important result is a discovery that the embryonic mechanism does allow not only super responses to the applied field but also makes these responses practically anhysteretic.

## Methods

### 1. Phase Field Microelasticity

The computational experiments of nano-embryonic thermoelastic equilibrium in pre-transitional austenite were carried out by using PFM theory and modeling of displacive transformations.[26-28] In this approach, the formation and evolution of MEs are described by the temporal and spatial distribution of the martensitic eigenstrain, $\varepsilon_{ij}^0(\mathbf{r})$, that is presented in terms of the long-range order (lro) parameters:

$$\varepsilon_{ij}^0 = \sum_{p=1}^{N} \varepsilon_{ij}^B(p)\eta_p(\mathbf{r}), \tag{1}$$

where $\varepsilon_{ij}^B$ is the Bain strain tensor describing the lattice misfit between the austenite and martensite in the stress-free state, and $\eta_p(\mathbf{r})$ $at$ $(p = 1 \ldots N)$ are *lro* parameters describing crystallographically equivalent orientation variants of the martensitic phase at point *r*, index *p* numbers these variants, N is the total number of possible Bain orientation variants. In the case of fcc→bcc MT, there are three orientation variants (N=3) with the Bain strains:

$$\varepsilon_{ij}^B(1) = \begin{pmatrix} \varepsilon_c & 0 & 0 \\ 0 & \varepsilon_a & 0 \\ 0 & 0 & \varepsilon_a \end{pmatrix}, \varepsilon_{ij}^B(2) = \begin{pmatrix} \varepsilon_a & 0 & 0 \\ 0 & \varepsilon_c & 0 \\ 0 & 0 & \varepsilon_a \end{pmatrix}, \varepsilon_{ij}^B(3) = \begin{pmatrix} \varepsilon_a & 0 & 0 \\ 0 & \varepsilon_a & 0 \\ 0 & 0 & \varepsilon_c \end{pmatrix}, \tag{2}$$

where $\varepsilon_c = a_{bcc}/a_{fcc} - 1$ and $\varepsilon_a = \sqrt{2}a_{bcc}/a_{fcc} - 1$ are the tetragonal transformation strains along the *c* and *a* axes and $a_{fcc}$ and $a_{bcc}$ are the lattice spacing of parent austenite and product martensite phases.

Since we consider the heterogonous pre-transitional austenite with static defects, the total stress-free strain entering the PFM model includes contributions from MEs and static defects and thus is a sum:

$$\varepsilon_{ij}^T(\mathbf{r}) = \varepsilon_{ij}^0(\mathbf{r}) + \varepsilon_{ij}^d(q)\chi_q(\mathbf{r}), \tag{3}$$

where $\chi_q(\mathbf{r})$ describes the spatial configuration of static defects, and $\chi_q(\mathbf{r}) = 1$ is inside static defects, dislocation loops or coherent nano-precipitates, otherwise $\chi_q(\mathbf{r}) = 0$.

The stress-free strain introduced by a dislocation in fcc crystals is:[29]

$$\varepsilon_{ij}^d(q) = \frac{1}{2d_{(111)}}(\mathbf{b}_{st} \otimes \mathbf{n}_t + \mathbf{n}_t \otimes \mathbf{b}_{st}), \quad q=(s,t)=1\ldots12, \tag{4}$$

where plastic deformation modes are $\langle 1\bar{1}0\rangle\{111\}$ dislocations in four $\{111\}$ planes, $\mathbf{b}_{st} = \frac{1}{2}a_{fcc}<1\bar{1}0>$ are the Burgers vectors in one of the three $<1\bar{1}0>$ directions ($s=1\ldots3$) in each of the four $\{111\}$ slip planes numbered by index t=1,…4, $\mathbf{n}_t$ is a normal to slip planes $\{111\}$, $d_{(111)} = \frac{1}{\sqrt{3}}a$ is the inter-planar distance, $a_{fcc}$ is the fcc lattice spacing, and q=(s,t) numbers 12 modes of the plastic deformation describing eigenstrains of all dislocation modes.

The simplest case of nano-precipitates that we considered is the one in which the misfit of precipitates is characterized by a dilatational eigenstrain:

$$\varepsilon_{ij}^d(1) = \begin{pmatrix} \varepsilon_p & 0 & 0 \\ 0 & \varepsilon_p & 0 \\ 0 & 0 & \varepsilon_p \end{pmatrix}, \tag{5}$$

where $\varepsilon_p$ is the misfit strain between the precipitating phase and the matrix which determine the potency of the defects.

The total free energy of the system was presented as a functional of $\eta_p$ as,

$$F = \int_V \left[ f_L(\eta_p) + f_G(\frac{\partial \eta_p}{\partial r}) \right] d^3r + E_{el}(\eta_p), \tag{6}$$

where $f_L$ and $f_G$ are the Landau free energy describing the bulk chemical properties and the gradient energy characterizing the contribution of the structural inhomogeneities to the chemical free energy caused by the local MT. $E_{el}$ is the total elastic strain energy produced by a given spatial distribution of $\eta_p$. The Landau free energy was approximated by a fourth-order polynomial meeting symmetry requirements of the parent phase,

$$f_L = \frac{1}{2} a_2 (\eta_1^2 + \eta_2^2 + \eta_3^2) + \frac{1}{3} a_3 (\eta_1^3 + \eta_2^3 + \eta_3^3) + \frac{1}{4} a_4 (\eta_1^2 + \eta_2^2 + \eta_3^2)^2, \tag{7}$$

The simplest form of the gradient free energy was chosen as:

$$f_G = \frac{g_2}{2} \sum_{p=1}^{3} \left[ \left(\frac{\partial \eta_p}{\partial x_1}\right)^2 + \left(\frac{\partial \eta_p}{\partial x_2}\right)^2 + \left(\frac{\partial \eta_p}{\partial x_3}\right)^2 \right], \tag{8}$$

where $x_1$, $x_2$, $x_3$ are coordinates aligned along the crystallographic <100> axes of the parent fcc phase. In the linear elasticity, the elastic energy can be presented as,[24-28]

$$E_{el}(\eta_p) = \frac{1}{2} \int_{k \neq 0} \frac{d^3k}{(2\pi)^3} B_{ijkl}(\mathbf{e}) \tilde{\varepsilon}_{ij}^T(\mathbf{k}) \tilde{\varepsilon}_{kl}^{T*}(\mathbf{k}), \tag{9}$$

where $B_{ijkl}(\mathbf{e}) = C_{ijkl} - e_m C_{ijmn} \Omega_{np}(\mathbf{e}) C_{klpq} e_q$, $\mathbf{e} = \mathbf{k}/k$ is a unit directional vector in the Fourier space, $\Omega(\mathbf{e}) = (C_{ijkl} e_i e_l)^{-1}$ is a Green function tensor and $C_{ijkl}$ is elastic modulus tensor. The super asterisk, *, designates the complex conjugate. $\tilde{\varepsilon}_{ij}^T(\mathbf{k})$ is the Fourier transform of $\varepsilon_{ij}^T(\mathbf{r})$ given by Eq.(3).

If an external stress is applied to the system, we have to add to Eq.(6) the coupling energy of the external stress with the MEs:

$$E_{el}^{ext} = -\sigma_{ij}^{app} \int_V \varepsilon_{ij}^T(\mathbf{r}) d^3r, \tag{10}$$

The energy term (10) describes a driving force for the MEs evolution provided by the applied stress. The evolution of the *lro* parameters $\eta_p$ was described by the Landau-Ginzburg equation as,

$$\frac{\partial \eta_p(\mathbf{r},t)}{\partial t} = -L_0 \frac{\delta F}{\delta \eta_p(\mathbf{r},t)} \tag{11}$$

where $L_0$ is the kinetic coefficient of the structure relaxation, and a variational derivative of the free energy functional $F$, $\delta F/\delta \eta_p(\mathbf{r},t)$, is a local transformation driving force at the point **r**.

We considered a generic example of a highly heterogeneous Fe-31at.%Ni austenite above the $M_s$ temperature ~ $239K$.[30] It has fcc crystal lattice with spacing $a_{fcc} = 3.58\text{Å}$. In the stress-free state, martensite has the bcc lattice with the lattice parameter $a_{bcc} = 2.87\text{Å}$.[30,31] Therefore, the components of the Bain eigenstrain of the martensite in Eq. (2) are $\varepsilon_a = 0.1322$, $\varepsilon_c = -0.1994$.

For simplicity, we assumed an isotropic elasticity with shear modulus $G_0 = 28GPa$ and Poisson ratio $\nu = 0.375$.[30] We assumed as in ref.[28] that $T_0 \approx 405K$ and $Q = 3.5 \times 10^8 \, Jm^{-3}$, where $T_0$ is the temperature of congruent stress-free equilibrium between the austenite and martensite, and Q is the latent heat related to the chemical driving force by equation,

$$\Delta \hat{f}_L = \hat{f}_L^{Aus} - \hat{f}_L^M = \frac{Q(T_0 - T)}{E_0 T_0}, \tag{12}$$

where $\hat{f}_L^{Aus}$ and $\hat{f}_L^M$ are the reduced chemical free energy of the austenite and martensite, respectively.

It is convenient to use reduced variables in numerical simulations. As in ref.[28], we introduced the computational grid size $l_0$ and typical energy of $E_0 = G_0(\varepsilon_c - \varepsilon_a)^2 = 3.0788 \times 10^9 \, Jm^{-3}$ as the length and energy unit. Then the reduced variables and parameters are:

$$\hat{x}_p = x_p/l_0 \; (p=1\ldots3), \; \hat{a}_s = a_s/E_0 \; (s=2\ldots4), \; \hat{g}_2 = g_2/(E_0 l_0^2), \tag{13}$$

Following ref.[28], we assume $\hat{a}_2 = 0.312$ is not dependent on temperature, $\hat{a}_3$ and $\hat{a}_4$ are linear function of temperature,

$$\hat{a}_3 = -0.936 + 1.36485(T-T_0)/T_0, \; \hat{a}_4 = 0.624 + 1.36485 \times (T_0 - T)/T_0 \tag{14}$$

Other parameters used in the simulations, e.g., $\hat{g}_2 = 0.01624$ and $\Delta \tau = 0.125$, are also chosen the same as in ref.[28].

## 2. "Sample" preparations

Since we studied pre-transitional austenite with the stress-generating defects, we had to start with a "preparation" of our "samples" of such austenite. To make the model more realistic, we did not use a priori assumptions about spatial configurations of static defects in these "samples". The distributions of dislocations or nano-precipitates in the initial defect-free austenite were obtained by solving the PFM kinetic equations of the dislocation dynamics and diffusional decomposition that describe a spontaneous self-organization of these defects. We employed PFM theory of dislocations,[27,32] which, in fact, is a dynamic generalization of the Peierls-Nabarro 1D static theory[29] of a dislocation line to the 3D case. In the PFM theory, the dislocation structure is not postulated. It is obtained as a result of spontaneous 3D evolution driven by the applied stress. The dislocation structure is spontaneously developed by a creation, annihilation and evolution of multiple interacting dislocations. In our simulation of the initial structure of the dislocated martensite, the "samples" with dislocation configurations consisting of perfect dislocations in all possible {111} slip planes along the <110> slip directions of the fcc crystal lattice were generated by simulating a plastic deformation stated by the operation of randomly distributed Frank-Read sources. The details of the PFM simulation of dislocation-induced plasticity can be found in refs.[27] and [32].

We introduced 80 randomly placed parallelogram plates with one grid in thickness as the Frank-Read sources in the fcc system. The types and sizes of all sources are also randomly chosen. Then we applied a uniaxial stress to the system along the [100] direction. When the plastic strain reaches about 35%, the applied stress is lifted, and the system containing complex dislocation configuration is allowed to relax towards the energy minimizing state in an absence of the applied stress. For maintaining of an unambiguous relation between the plastic strain and the dislocation density, only samples with no dislocation loops swept through the simulation box were kept for further investigation. During the relaxation dislocation structures with specific amounts of plastic strain, ranging from about 0.1% to 30%, were chosen as the static initial state for further simulations of MEs formation. The obtained typical dislocation microstructures of the initial state of the pre-transitional austenite are shown in Fig. 1.

"Samples" of the initial pre-transitional austenite with coherent nano-precipitates as static defects were generated by prototyping the early stage of decomposition by using the PFM theory of decomposition.[33,34] The total free energy of the decomposing pre-transitional solid solution was chosen as a functional of concentration field c(**r**):

$$F = \int [f_c(c) + \frac{\beta}{2}(\nabla c)^2] d^3r, \qquad (15)$$

where the gradient term describes the interfacial energy with $\beta$ being a coefficient, and $f_c(c)$ is the chemical free energy. We used the simplest double-well polynomial,[33,34]

$$f_c(c) = A(c-c_0)^2(c-c_1)^2, \qquad (16)$$

where A is the energy constant of the solid solution, $c_0$ and $c_1$ are the equilibrium concentrations for two product phases. The chosen parameters for the decomposition are:

$$c_0 = 0.05, c_1 = 0.095, A = 15.6, \beta = 7.8, \qquad (17)$$

and a typical initial concentration of $c = 0.5$ is used for simulating of the decomposition. According to Eq.(3) with q=1, the structural inhomogeneity is defined as a functional of composition,

$$\chi_1(\mathbf{r}) = \frac{c(\mathbf{r}) - c_0}{c_1 - c_0}, \qquad (18)$$

Since the isostructural coherent decomposition in an elastically isotropic solid solution has no effect to the formation of microstructure,[24] the elastic energy is omitted during the preparation of samples.

## Results

### 1. Effects of Thermal Cycling

We first simulated the change of MEs in pre-transitional austenites with different amounts of plastic strains under the cycling of temperature (no external stress is applied). Typical microstructures of the sample with ~5% plastic strain at different temperatures above the $M_s$ are shown in Fig.2. The $M_s$ temperature for samples with different plastic deformations were estimated as temperatures at which the defect-bounded MEs lose their stability and start to grow, transforming all austenite into martensite, Fig. 3. It is seen that the estimated values of $M_s$ strongly depend on density of defects. In particular, $M_s$ monotonously increases upon increase of dislocation density.

The temperature-induced expansions or contractions with different densities of dislocations are simulated at different temperatures above $M_s$. The results, summarized in Fig. 4, show how changes of temperature influence the dislocation-induced embryos in samples with different plastic deformation: the nano-scale equilibrium size of embryos and their equilibrium volume fraction increase upon cooling towards the $M_s$ temperature and reversibly decrease upon heating.

Figure 4 thus demonstrates that the ME-induced macroscopic strain describing a thermal expansion upon cooling is an effect that is opposite to the conventional thermal contraction upon cooling caused by anharmonicity of atomic vibrations, and thus a contribution of the MEs to the thermal expansion coefficient is negative. Figure 4 shows that the simulated dependence of the strain generated by MEs on temperature is practically non-hysteretic upon thermal cycling. Therefore, measurements of the thermal extension/contraction, which is a sum of the conventional anhysteretic positive thermal expansion and the negative one caused by MEs, cannot single them out. If the negative contribution to the thermal expansion coefficient by MEs just cancels the conventional positive contribution caused by anharmonicity of atomic vibration, the thermal expansion vanishes. This vanishing of the thermal expansion is called Invar effect.[35-37] It is interesting that the classical Invar alloys are pre-transitional austenites with the fcc crystal lattice. That is, the fcc→bcc MT in these alloys are really observed at low temperature, e.g., $M_s$ = -223°C for Fe-33at%Ni.[38]

## 2. ME-Induced Strain Responses to Applied Stress

The strain response to the applied stress is associated with the stress-induced changes of the sizes and volume fraction of MEs. In the simulation, the applied external stress, $\sigma_{ij}^{app}$, was chosen as a pure shear stress:

$$\sigma_{ij}^{app} = \frac{\sigma}{2}\begin{pmatrix} 0 & 0 & 0 \\ 0 & 1 & 0 \\ 0 & 0 & -1 \end{pmatrix}, \quad (19)$$

where the tensor $\sigma_{ij}^{app}$ is presented in the Cartesian coordinate system whose axes are directed along the <100> axes of the fcc austenite lattice. According to the Hooke law, this stress induces shear strain in the (011) plane along the $[0\bar{1}1]$ direction. In this case, the shear modulus in the Hooke's law is $C' = (C_{11} - C_{12})/2$.

The applied shear stress was chosen in the form (19) because there are numerous studies reporting a reduction of $C'$ upon cooling prior to the MT.[5,6,39] This elastic softening is often interpreted as intrinsic phenomenon that is also the reason for the underlying MT. However, as shown in our simulation, the MT can develop even without the softening of $C'$ because the observed softening can be just an extrinsic effect caused by the ME-producing defects.

Indeed, the applied stress, $\sigma_{ij}^{app}$, produces two effects, viz., the conventional Hookean strain, $\bar{\varepsilon}_{sh}^{H} = \sigma/C'$, where $C'$ is the intrinsic shear modulus of the defect-free austenite and an additional embryo-induced strain, $\bar{\varepsilon}_{sh}^{ME}(\sigma)$, that is generated by the MEs and modified by the applied stress. The additional strain, $\bar{\varepsilon}_{sh}^{ME}(\sigma)$, produced by MEs can make giant the total strain, $\bar{\varepsilon}_{sh}^{T} = \bar{\varepsilon}_{sh}^{H} + \bar{\varepsilon}_{sh}^{ME}$, with respect to the conventional Hookean strain, $\bar{\varepsilon}_{sh}^{H}$. This is the strain amplification occurring at the same applied stress that results in the elastic softening.

Since the rate of a MT producing ME is practically always faster by orders of magnitude than the rate of externally applied stress, the evolution of MEs is so fast that it provides an attainment of equilibrium configurations at each instantaneous value of the time-dependent stress. The latter indicates that a martensitic response to the cyclically applied stress is a quasi-static process and the values of the response strain are equilibrium values. If the strain response is also giant, which is generally the case, the defected austenite in the pre-transitional state is superelastic. Our modeling of the stress response is carried out to confirm the idea that an engineering of a defected state of a pre-transitional austenite by plastic deformation and/or aging can be used as a general approach to design superelastic alloys and Invar alloys. It is practically not important whether MEs are generated by dislocations or coherent precipitates. What is important is that the defects can generate stress sufficient to promote the formation of MEs.

In our simulation, we modeled a quasi-static microstructure evolution by slowly applying the time-dependent stress (19) to dislocated pre-transitional austenite at $T/M_s \sim 1.15$. This stress field can result in either the growth/shrinkage of already existing MEs or the formation of MEs around dislocations that were absent without this stress. It is noted that this process can be interpreted as a field-induced isothermal MT in spatially inhomogeneous stress field that is localized in the vicinity of embryo-assisting dislocations. Typical microstructures of MEs at different values of applied stress in a sample with about 5% plastic strain are shown in Fig.5. Fig.5a demonstrates that MEs already exist above the $M_s$ temperature even without applied stress. With increasing the stress level, the "blue" embryos shrink and the "green" ones grow both along the dislocation loops, where colors distinguish different orientation variants of MEs Figs.5b and 5c. A reversal of the stress results in the reversal of the ME structure along the dislocation lines, Figs. 5e and 5f. It can be seen that the microstructures at the same loading and unloading stress are almost the same (Figs.5b and 5f, Figs.5c and 5e). This indicates that the growth/shrinkage of embryos is just weakly path-dependent, and thus the macroscopic responses are low-hysteretic.

The nonlinear macroscopic strain response to the applied stress is shown in Fig. 6. It has been noted that the total strain, $\bar{\varepsilon}_{sh}^{T}$, consists of two terms: the conventional Hookean strain of the embryo-free austenite, $\bar{\varepsilon}_{sh}^{H}$, and an additional embryo-induced strain, $\bar{\varepsilon}_{sh}^{ME}$, generated by an appearance and/or change of MEs. Figure 6 shows a trend demonstrating that a sample with higher density of dislocations has larger and more sharply increasing embryo-induced strain, and this strain is recoverable with weak hysteresis.

The softening of shear modulus caused by the extrinsic contribution of MEs is demonstrated in Fig. 7, where the effective shear modulus $C'_e$ is defined as $C'_e = \sigma_{ij}^{app}/\bar{\varepsilon}_{sh}^{T} = \sigma/(\bar{\varepsilon}_{22}^{T} - \bar{\varepsilon}_{33}^{T})$. It is shown that the isothermal modulus softening (decrease of shear modulus of the defected austenite, $C'_e$), become more significant with an increase of the applied stress and dislocation density.

The shear modulus softening upon cooling was also investigated at two fixed values of stress, Fig. 8. The dotted lines in Fig. 8, are guides to eyes to show the lowest temperature ($M_s$) at which the defect-bounded embryos still exist at a given stress level. It can thus be seen that under both stress levels, all samples show similar softening upon lowering the temperature towards their $M_s$. Meanwhile, the moduli of samples with higher densities of dislocation were softened gradually but significantly over a relatively wide temperature range. For example, the smallest value of the effective shear modulus, $C'_e$, obtained for the sample with ~30% plastic strain reached a value of ~10% of $C'$ over a temperature range of ~100K, whereas that for the sample with ~0.1% plastic strain reached just ~0.9 of $C'$, Fig.8b.

We also investigated the effects of nano-precipitates on the formation and evolution of MEs in the pre-transitional austenite. In this case the stress-generating defects are precipitates. To compare the effect of dislocations and coherent precipitates, we used the same free energy functional in which only the stress energy contribution of defects was modified. The defect structure formed by nano-precipitates was simulated by prototyping the early stage of decomposition by using the PFM kinetic equation of decomposition described in the model section. Since a potency of nano-precipitates promoting the MEs formation is usually much lower than that of full dislocations, a high density of precipitates (~50%) was used. Figure 9 showed typical MEs microstructures at different values of applied stress, where green visualizes the static nano-precipitates and blue are MEs. The MEs can grow/shrink upon increase/decrease of the applied stress. It was also shown in Fig.10 that the macroscopic response to applied stress is either low-hysteresis or anhysteretic. This response is similar to the effect produced by dislocations. The shear modulus of samples also become softer when the temperature is lower and the potency of heterogeneities (determined by the dilatational misfits, $\varepsilon_p$) is higher, Fig.10.

3. **ME-Induced Strain Responses to Applied Magnetic Field**

The simulated mechanism of superelasticity caused by the shift of thermoelastic equilibrium between the MEs and pre-transitional defected austenite above $M_s$, actually, provides a much more general concept of utilizing MEs to design materials with multi-ferroic super response. In particular, these materials can be super-magnetostrictive pre-transitional austenitic alloys. The latter is possible if the martensitic phase is ferromagnetic whereas austenite is either paramagnetic or antiferromagnetic or vice versa. In principle, both martensite and austenite phases could be ferromagnetic but they need to have different saturation magnetizations.

Fortunately, there is no principal difference between the ME-induced strain super responses to the applied stress and to the applied magnetic field.[40,41] The difference is only in the physical nature of the driving force. The formation of MEs near defects and their responses to different driving forces are conceptually the same. In this case the applied magnetic field plays the same role as the applied stress. The results obtained for superelasticity in the FeNi alloys can thus be extended to predict a possibility of supermagnetostriction of a defected pre-transitional austenite because the austenite of FeNi alloys is paramagnetic and martensite ferromagnetic. If such an embryo-related giant strain is achieved in pre-transitional austenite, this material would be a generic prototype of a completely new class of supermagnetostrictive materials competitive to Terfenol-D.[4] This material, as a rule, could be much cheaper and potentially could have a stronger magnetostriction.

To simplify the model, we assumed a strong magnetocrystalline anisotropy with the <100> directions of easy magnetization. Under this assumption, the spontaneous magnetizations of the orientation variants are rigidly bounded to the easy magnetization direction of their orientation variant. In other words, this approximation neglects deviations of saturation magnetizations from their easy magnetization directions. All changes of the magnetic state are provided by configurational changes of the martensitic orientation variants. This assumption not only significantly simplifies both theory and computer simulation of the magnetostriction but even makes the new simulations unnecessary if the magnetic field is applied along the $[0\bar{1}1]$ direction. It turns out that the reduced form of the PFM kinetic equations used for superelasticity is the same as for the supermagnetostriction. To emphasize the analogy between the superelastic and supermagnetostrictive response, we introduce a concept of an equivalent "magnetic stress", $\sigma^{mag}(H)$, that is defined as,

$$\sigma^{mag}(H) = \frac{2\mu_0 M_0}{\varepsilon_c - \varepsilon_a} H, \tag{20}$$

where $\mu_0$ is the vacuum permeability of a single domain martensite, $H$ is the applied magnetic field and $M_0$ is a saturation magnetization, $\varepsilon_c$ and $\varepsilon_a$ are diagonal components of the martensitic eigenstrain, respectively. By introducing the typical magnetic field defined as $H_o = C'(\varepsilon_c - \varepsilon_a)/2\mu_0 M_0$ and using it as a scaling factor, a magnetic field can be related to a stress as $\sigma^{mag}/C' = H/H_0$. With this definition, all our results obtained for the strain response to the applied stress can be used for the strain response to the applied magnetic field. We just have to use everywhere the magnetic stress $\sigma^{mag}(H)$ instead of mechanical stress. Therefore, we do not need to carry out separate computer simulations of the strain response to the magnetic field. All previous plots describing this response are still valid after this substitution. In this sense, *supermagnetostriction is actually a superelasticity caused by an application of equivalent magnetic stress, $\sigma^{mag}(H)$*.

In this paper, typical values for Fe-31at%Ni alloy ( $\mu_0 = 1.26 \times 10^{-6}$ N/A² and $M_0 = 1.46 \times 10^6 A/m$, interpreted from magnetizations of iron and nickel according to the atomic fraction) were used to estimate $H_0 \sim 2.5 \times 10^9$ A/m ( $\square\, 3.1 \times 10^4 kOe$ ). With this estimate, the magnetic-strain curves are shown in Fig.6a and Fig.10a. Therefore, if such magnetic materials are in a pre-transitional state, the responses of defect-induced pre-existing ferromagnetic MEs to the magnetic field could provide the desired supermagnetostriction. It is noted that the approximation of strong magnetocrystalline anisotropy, in principle, is not critical. A general case of the magnetic anisotropy just requires taking into account deviations of the magnetizations from the easy magnetization directions. This increases the number of phase field variables but still makes computations tractable. It would not "derail" the simulation of the supermagnetostriction, but just make them more expensive.

## Discussion

Electron microscopic observation of precursor states consisting of nano-size structural clusters that presumably are MEs has a long history.[7] This structure often has a tweed-like patterned,[5-7,13] which was usually observed in pre-transitional states of the austenite and order-disorder transitions. The previous 2D computational modeling demonstrated that the nano-structured mixed states are observed in a presence of randomly distributed point defects.[19-21] Actually, without the presence of these defects the tweed-like structure is unstable and gradually transforms into a conventional polytwinned stress-accommodating structure.

Using the 3D PFM computer simulation, we demonstrated that defects (dislocations and nano-precipitates) generating a significant local stress do induce the stable nano-MEs in the otherwise stable pre-transitional austenite. The production of MEs is actually the localized stress-induced MT. Such a process would be thermodynamically impossible in the defect-free austenite because it increases the free energy. However, the local stress generated by defects decreases the energy cost of formation of the MEs near the defects and thus allows them to grow until the size of MEs reaches the value providing the thermoelastic equilibrium. In this situation, the total volume of MEs and their density, which are commensurate with density of static defects, are thermodynamic parameters varying with the temperature and external field and eventually assuming the equilibrium values.

Our simulation demonstrated that the relation between the total volume of MEs and temperature critically depends on how far is temperature of the system, T, from the $M_s$ temperature. The closer the temperature $T$ to $M_s$, the greater the total equilibrium volume of MEs, i.e., cooling always increases the volume fraction of MEs (Fig.4). Our results clarify the physical meaning of the $M_s$ as an instability temperature at which the MEs "detach" from the static defects and start to expand until they absorb the entire austenite. Therefore, $M_s$ is an extrinsic characteristic determined by the potency and density of defects.

The thermoelastic equilibrium of the pre-transitional state obtained in this paper can also explain the origin of the diffuse phase transformation (DPT) that refers to a gradual, reverse and anhysteretic isothermal displacive phase transformation upon cooling and under applied field, although the DPT was observed and best studied in ferroelectrics.[42-44] Actually, the existence of DPT directly follows from the thermodynamics of nano-embryonic thermoelastic equilibrium: our simulations predict that the shift of thermoelastic equilibrium gradually increases the total volume of MEs upon cooling to the $M_s$/Curie temperature, and thus gradually increases the ME-induced strain/polarization/magnetization, Fig.4. For example, the neutron diffraction study of lead magnesium niobate (PMN) and Ta-bearing strontium barium niobate (SBNT)[44] demonstrated that the volume of polar microregions within paraelectric matrix (these microregions are ferroelectric phase embryos formed by displacive ferroelectric transformation) and dielectric permittivity gradually increases upon cooling towards the Curie temperature.

Our results also provide a new extrinsic ME-related mechanism to explain the Invar effect observed in FeNi and many other alloys. The previous attempts to explain the Invar effect by magnetic phenomena [36,37] are not satisfactory since they cannot explain the fact that the Invar effect is also observed in non-magnetic alloys.[5] However, the Invar effect can also follow from the thermodynamics of pre-transitional austenite with stress-generating defects. Indeed, as is well-known,[23,24] the formation of coherent inclusions of a new phase (MEs in our case) can produce a macroscopic strain changing the shape and volume of a sample. If the volumetric effect of the austenite→martensite transformation is positive, as it is in the considered Fe-Ni alloys, then, as follows from our simulations, the homogeneous strain generated by the growing MEs increases the sample volume upon cooling towards $M_s$ and decreases it upon heating. This is a special thermal-volumetric effect that is opposite to the conventional thermal expansion/contraction. If these two opposite contributions cancel each other, the material has no thermal expansion and we would have an Invar effect. Therefore, the Invar effect is expected in pre-transitional defected austenite if the martensite has greater atomic volume than austenite (otherwise, the ME mechanism just amplifies the conventional effect of thermal contraction upon cooling). This nano-embryonic mechanism allows one to tune-up the Invar properties either by doping the austenite to vary its $M_s$ temperature and thus the chemical transformation driving force, or by plastic deformation controlling dislocation density, [45-48] or by aging to control the volume fraction and density of coherent precipitates.

Similarly, the ME mechanism may provide the so-called Elinvar effect[36], an independence of the elastic modules on temperature. It was demonstrated in Fig.8 that the growth/shrinkage of MEs under applied stress provide a contribution to the elastic softening. This effect increases upon cooling since the cooling increases the volume fraction of MEs responsible for the softening. The softening is an effect that is opposite to the conventional effect of hardening upon cooling caused by the anharmonic properties of the defect-free austenite.[5,49,50] If these two effects cancel each other, the elastic modules become independent on temperature, and we have the Elinvar effect.

Our results demonstrated that the giant responses of pre-transitional materials with stress-generating defects are either anhysteretic or at least low-hysteretic (an exception is the response in a close vicinity to $M_s$ temperature where some hysteresis sometimes appears). The anhysteretic behavior is associated with the fact that the formation of the stable nano-MEs in the defected pre-transitional austenite allows the system skip the nucleation stage of the MT. Fig.6 and Fig.10. This is a general conclusion applicable to a wide spectrum of systems because the behavior of all of them is dictated by the same thermodynamics of the pre-transitional state and thus is not system specific.

The discussed ME mechanism is applicable for engineering superelasticity of pre-transitional materials by sever plastic deformation producing a high density dislocations or by aging producing nano-precipitates. The deformation-produced superelasticity is supported by a discovery of the so-called Gum Metals.[35] This alloy is a bcc solid solution that, depending on composition, undergoes either the bcc→$\alpha''$ or bcc→$\omega$ MT at low temperatures.[51,52] After a severe plastic deformation (~90% reduction in area), the alloy acquires special characteristics like ~2.5% of fully anhysteretic nonlinear elastic strain response to the applied stress, and excellent Invar and Elinvar effects over a wide temperature range.[35] All of these superior properties could be well explained by the extrinsic ME mechanism, because the bcc solid solution in this case is a pre-transitional austenite with a MT at low temperature; the severe plastic deformation of austenite generates high density of dislocations; the observed coherent "nanodisturbances" with the new phase structure[16,35] can be interpreted as MEs. The doping by V,Ta,Nb,O at compositions close to the congruent boundary between the $\alpha''$ and $\omega$ martensites may significantly change the $M_s$, which, in turn, may change the driving force of MEs formation increasing the nonlinear superelasticity effect, as well as tuning up the Invar and Elinvar effects.

The second example is the superelasticity observed in NiTi after aging and plastic deformation[5] and NiTi:Fe alloys after aging.[15] The aging treatment also produces a dense distribution of martensitic nano-particles in NiTi:Fe,[15] which is very similar to the one observed in Gum Metals. For example, the high-temperature austenitic state in NiTi alloy near equiatomic stoichiometry is also a bcc solid solution. It is also a pre-transitional phase because it also transforms into two different martensitic phases (monoclinic B19' and rhombohedral R phases) at low temperature.[5,15]

The Fe-30at%Pd alloy with a pre-martensitic tweed-like nano-dispersion of single domain clusters of martensitic phase could be probably added to this list as well. This alloy has about 7-fold softening of the elastic modulus, $C'$, near but above the $M_s$.[6,12]

As follows from Eq.(20), supermagnetostriction is equivalent to the "superelasticity" caused by the "magnetic stress". Indeed, supermagnetostriction observed in some FSMAs seems to be also explained by the defect-induced MEs responding to the applied magnetic field. A particular example is the Fe-Ga alloys with two magnetostriction peaks of about 400ppm at compositions close to 19at%Ga and 24at%Ga[53,54] (for a comparison, the value for pure $\alpha$ Fe is ~10 ppm[55]). There are reasons to believe that the obtained magnetostriction is an extrinsic effect associated with ME mechanism rather than with the conventional spin-phonon interaction in a homogeneous phase, because (i) the magnetostriction strongly depends on its thermal history: a quenching of the alloy always produces much higher magnetostriction than a slow cooling;[56] (ii) the dependence of magnetostriction on composition has sharp peaks, and the locations of these peaks are not accidental—they coincide with the solubility limits of the disordered bcc and $DO_3$ phases on the phase diagram—at these compositions, the bcc and $DO_3$ ordered alloys start to decompose forming the stress-generating coherent precipitates; (iii) HRTEM images did confirm the existence of coherent nano-clusters of the precipitate phase at compositions of the peak of magnetostriction about 19 at %Ga;[53] (iv) the effect of superelasticity was really observed in the Fe-Ga alloy near 24at%Ga[54]—the observed peak of magnetostriction in this alloy could also be interpreted as a superelasticity caused by the equivalent magnetic stress.

Furthermore, a giant magnetostriction together with superelasticity were also observed in the Fe-Pd alloys with $M_s$ ~252K.[57,58] A significant enhancement of magnetostriction, which can reach ~1000ppm in a single phase state was also reported in the bcc Fe-Co alloys near the fcc/bcc phase solubility limit [59] where the formation of coherent nano-precipitates of cubic intermetallic by decomposition was observed.[60] Moreover, the study of $Fe_xCo_{1-x}$ alloys near solubility limit at x=0.25 showed that plastic deformation also increases the magnetostriction from 70 to 120ppm.[60] This is also expected if we assume the ME-mechanism predicting the magnetostriction increase with plastic deformation, Fig.6a, and coherent precipitation of intermetallic.

The concept of thermoelastic equilibrium in the pre-transitional state can also be extended even further to a scenario in which alloys has coherent nano-precipitates. However, the potency of these defects is not sufficient to produce MT but is nevertheless sufficient to produce them with an extra help from the applied field. In this case, the localized MT will only occur under applied field forming a mix state with MEs. The responses of these MEs to applied field will also be recoverable, anhysteretic, and giant. However, conventional HRTEM (without applied field) can only observe nano-clusters formed by the precipitation of the second phase. To observe MEs, the HRTEM observations should be performed under a properly applied stress or magnetic field.

It should be noted that *the presence of MT is not a necessary condition required to form the MEs*. Indeed, if the free energy of the martensite is close to but still higher than that of austenite at all temperatures, the MT does not develop at all. However, high density of high-potency stress-generating defects may still be sufficient to induce MEs. In this scenario, the system should have all properties obtained in this study, viz., the system should form the MEs in the thermoelastic equilibrium with the matrix, should have the diffuse MT (however, the diffuse transformation in this case may not necessarily end up by the conventional MT), and finally should have a recoverable and low-hysteretic/anhysteretic super-responses to applied fields.

Although in this paper we have only investigated the effects of defect-bounded embryos in austenitic systems which is stable in the defect-free state, the obtained insights can be naturally extended to another type of conceptually similar materials with morphortropic phase boundaries (MPBs). This is a situation where two structurally different phases formed from the same parent phase are separated by a boundary line of the congruent equilibrium. In this case, one of these phases can be regarded as pre-transitional austenite, and another as the martensite. Therefore, a MPB system with stress-generating defects (or electric-field-generating defects in cases of ferroelectric solid solutions) may also have giant anhysteretic strain response to the applied fields.

In summary, we have demonstrated that a presence of stress-generating defects in a pre-transitional austenite is responsible for the new phenomenon. This phenomenon is a thermoelastic equilibrium between the defect-induced nano-size MEs and the pre-transitional austenite whose defect-free homogeneous state is stable. The MEs are formed as a result of the localized stress-induced MT developing around defects. Varying the temperature and/or applying external field shifts the thermoelastic equilibrium by affecting the size and volume fraction of MEs. This causes a giant, recoverable and low-hysteretic/anhysteretic ME-induced strain response to the temperature and/or external fields. The latter is a new mechanism directly following from the general thermodynamics of pre-transitional alloys in the structurally heterogeneous state. The effect is generic since it can be expected in systems with different physical nature. In particular, the ME mechanism can explain superelasticity, supermagnetostriction, elastic softening, Invar and Elinvar effects and may provide a guidance to engineering of a new class of material with giant anhysteretic or low-hysteretic response to applied stimuli.

# References:

Email: *wfrao@nuist.edu.cn*

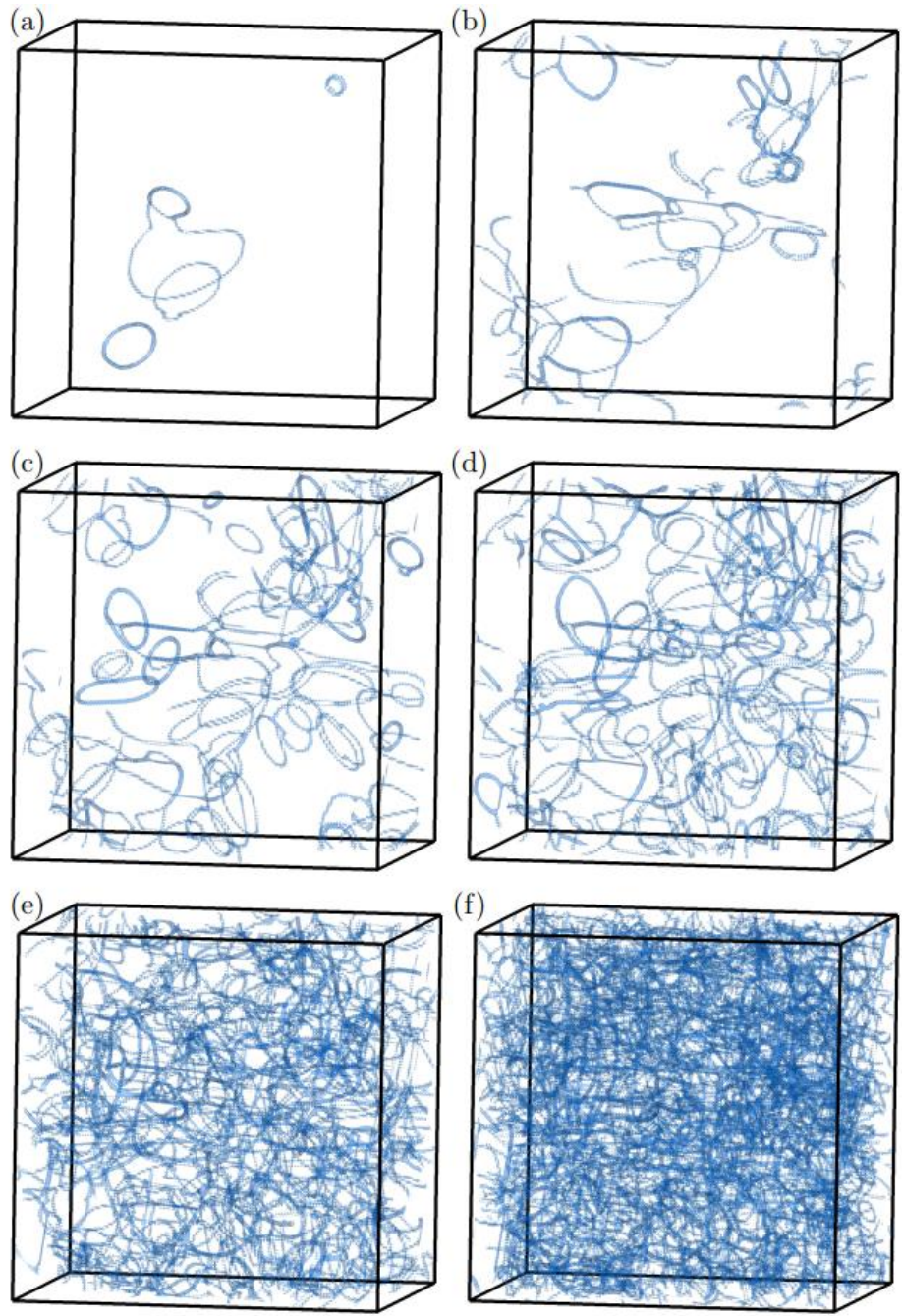

**Figure 1**: Simulated dislocation structures of the pre-transitional austenite prior to the formation of MEs. The plastic strains are: (a) 0.1%, (b) 0.5%, (c) 1%, (d)2%, (e)5% and (f)10%. The dislocations are generated from randomly placed Frank-Read sources. The applied stress is uniaxial along the [100]-direction. The microstructures with higher dislocation densities are not shown because of the difficulty of visualizing the individual dislocations.

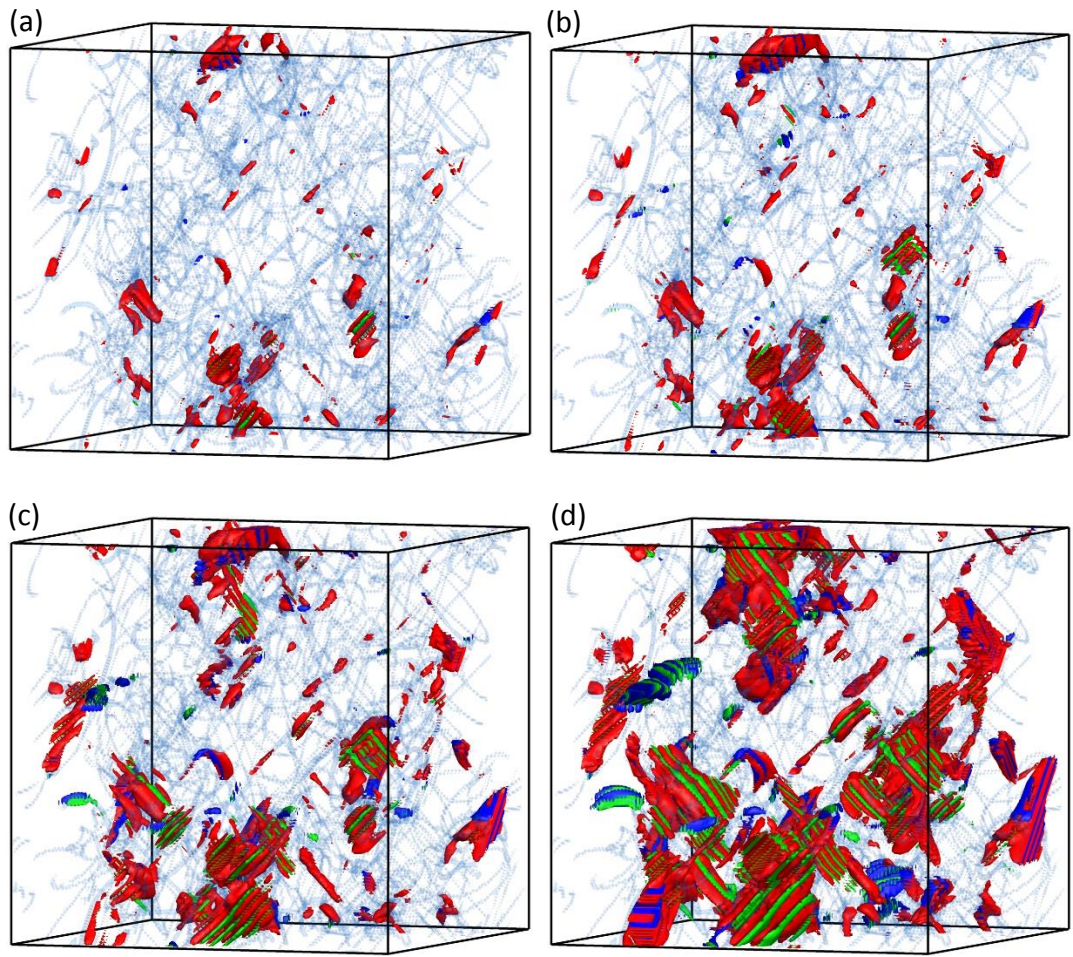

**Figure 2**: The simulated evolution of nano-embryonic martensitic structures in a dislocated pre-transitional austenite with ~5% plastic strain upon temperature cycling. The simulation is performed without applying external stress. Typical microstructures are shown at (a) T/$M_s$~1.2, (b) T/$M_s$~1.14, (c) T/$M_s$~1.07 and (d) T/$M_s$~1.01, where the dislocation lines are shown in cyan, and three orientation variants of MEs are shown in red, green and blue, respectively.

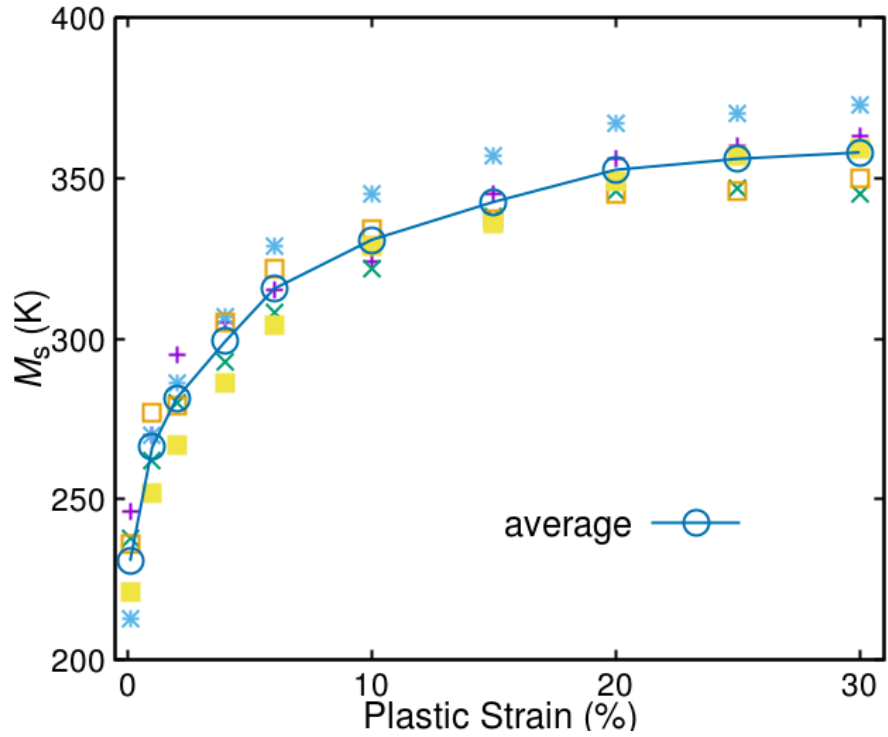

**Figure 3:** Simulated values of $M_s$ versus the previous plastic strain, where data points show the $M_s$ measured for five sets of different dislocation configurations and the average values of $M_s$ are connected by segments to guide to the eyes.

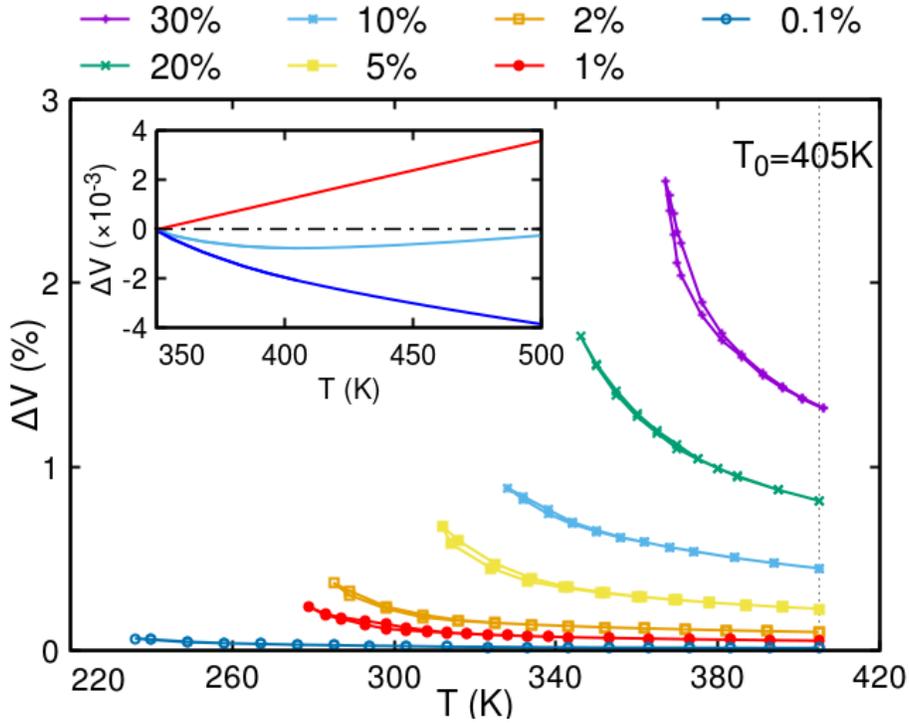

**Figure 4:** Simulated relative volume changes (thermal expansion/contraction) versus temperature for samples with different amounts of plastic strain. Inset: volume change versus temperature for an isotropic alloy with a volumetric coefficient of thermal expansion of $\alpha = -24 \times 10^{-6}/K$ [35] (red line), the simulated volume change versus temperature for a sample of ~10% plastic strain (blue line), and the sum of both (cyan line).

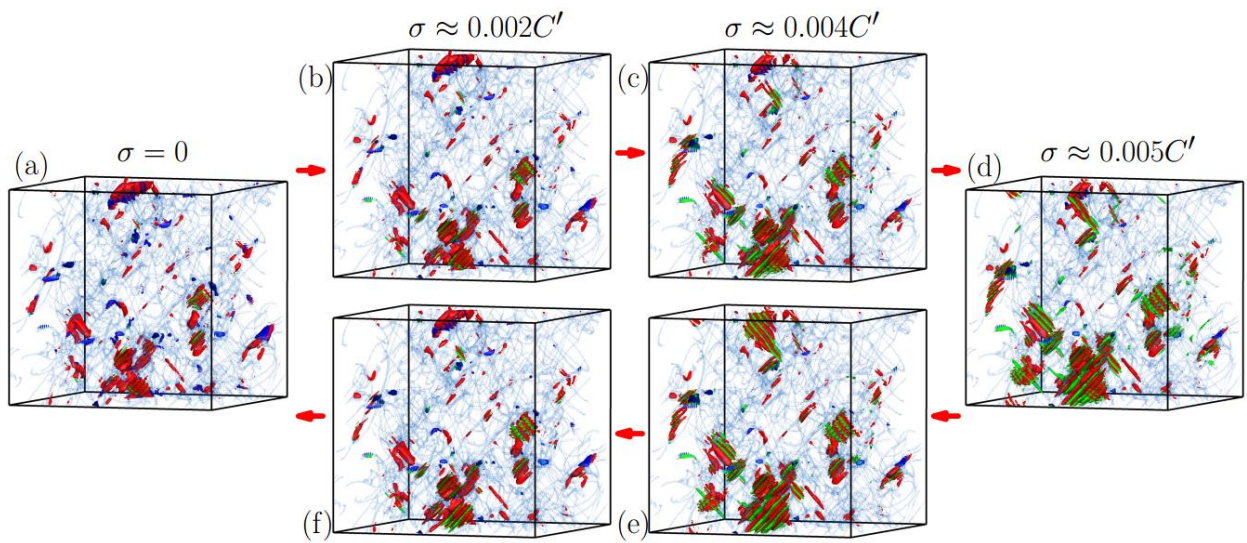

**Figure 5:** The simulated evolution cycle of nano-embryonic martensitic structures in the dislocated sample of pre-transitional austenite with ~5% plastic deformation caused by quasi-static application of external stress. The simulation is performed at T/$M_s$~1.15. The static dislocation structure visualized by dislocation lines was generated by a preliminary computer simulation of plastic deformation of defect-free austenite. The lines are shown in cyan. MEs of different orientation variants are colored in red, blue and green, respectively. The red arrows indicate the sequence of the structures obtained as the stress is cycled.

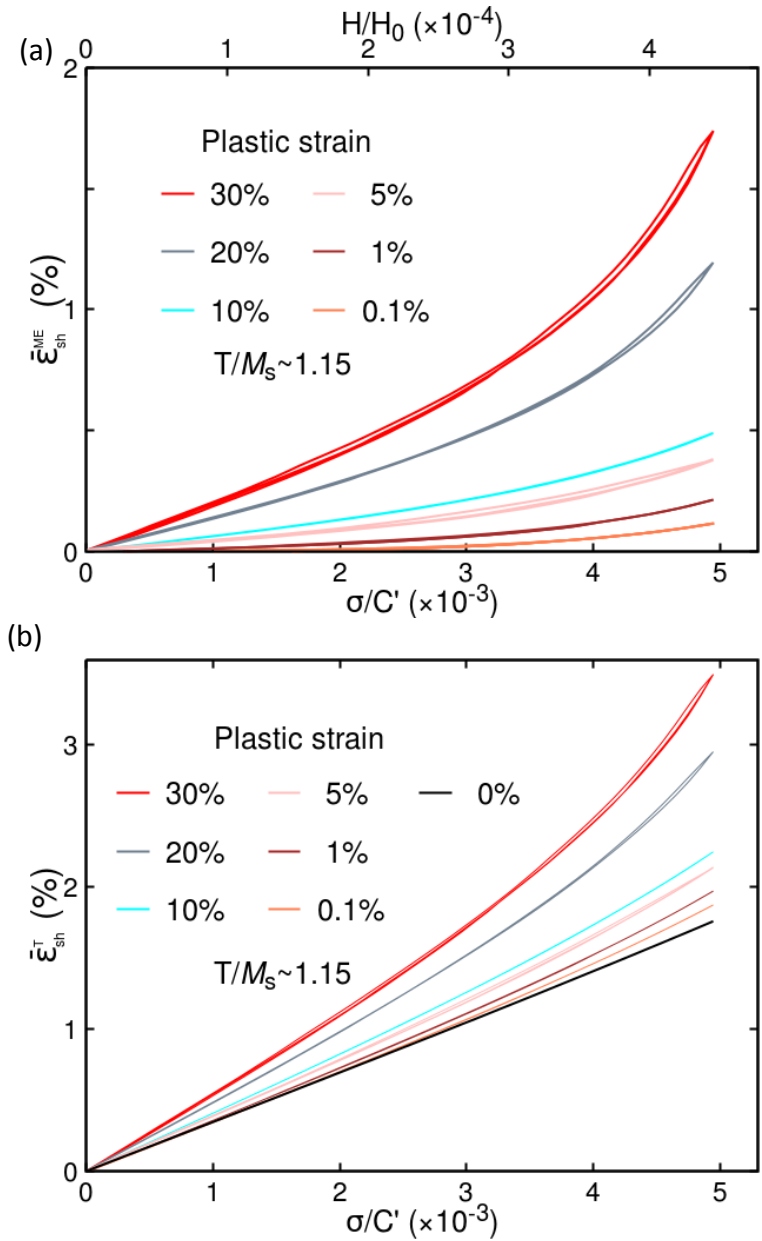

**Figure 6**: (a) Simulated shear strain responses to the applied fields provided only by the change of MEs, and (b) the total shear strain responses to a quasi-statically applied shear stress at reduced temperature of T/$M_s$~1.15 for pre-transitional austenites with different amounts of plastic deformation. In (a), the reduced magnitudes of the applied stress and magnetic field are shown on the lower and upper abscissas, respectively. In (b), the black line describes the Hookean strain response (without the contribution from the change of MEs). Note the responses are almost anhysteretic—the hysteresis loops are extremely slim.

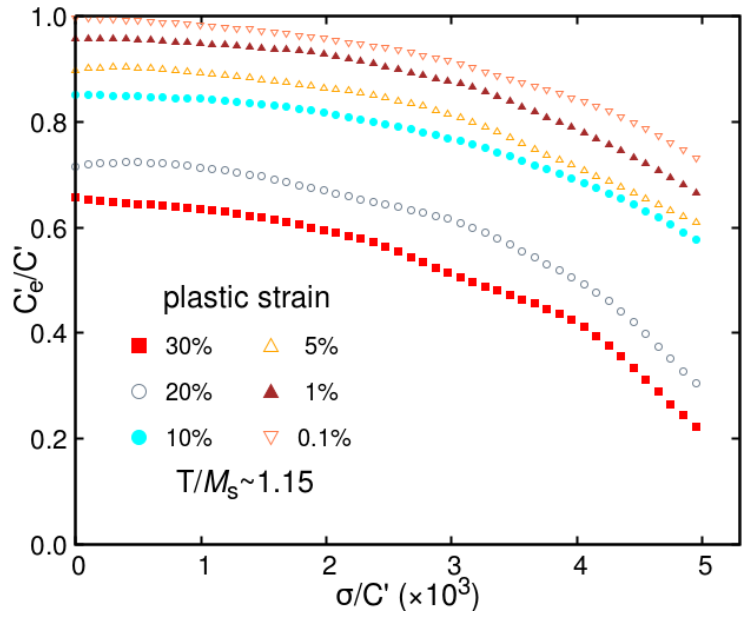

**Figure 7**: The simulated effective shear modulus versus the applied stress. The values are calculated from the lines shown in Fig.6b for samples with different density of dislocations at T/$M_s$~1.15.

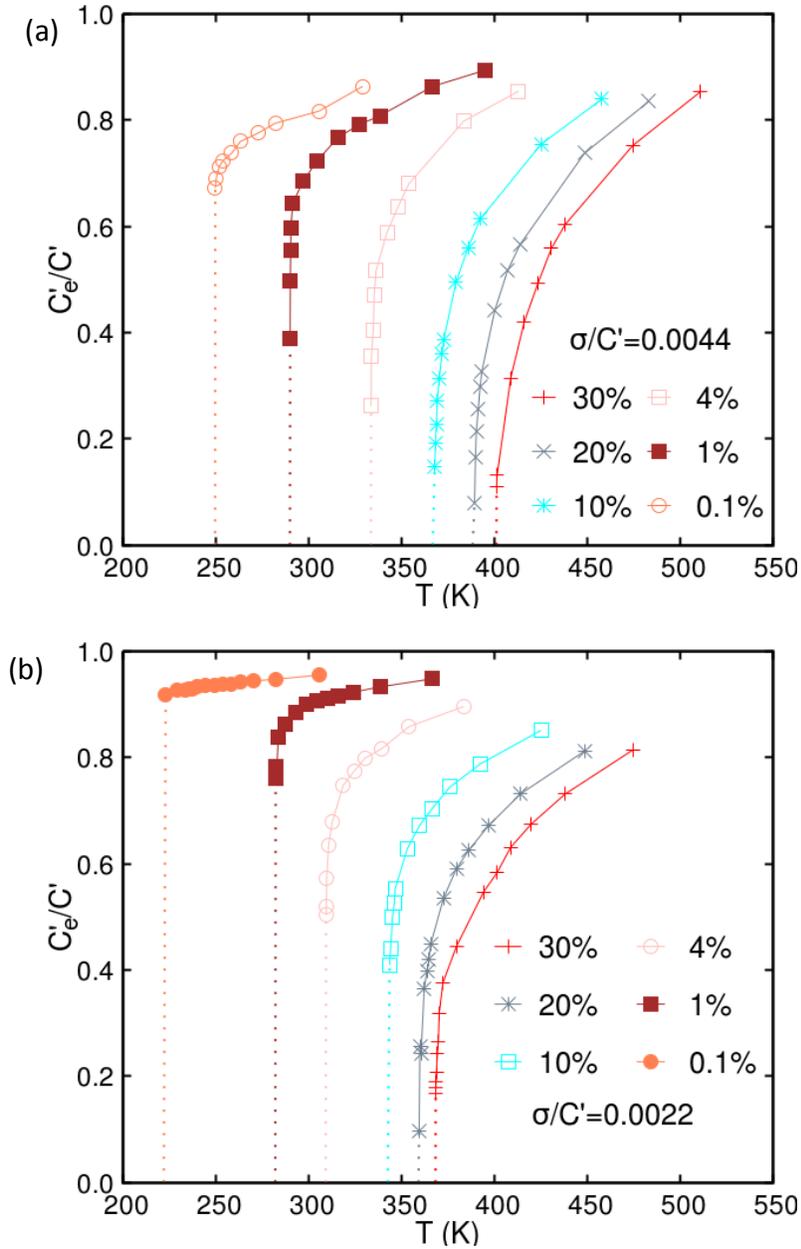

**Figure 8**: Simulated temperature dependence of the effective shear modulus for different amounts of plastic strain. The modulus was calculated at a constant shear stress level of (a) $\sigma/C' = 0.0044$ and (b) $\sigma/C' = 0.0022$, where $C'$ is the shear modulus of homogeneous austenite. Dashed lines are guide to the eyes showing the temperatures at which MEs grow through the whole simulation box. The figure illustrates a drastic softening of the shear modulus of pre-transitional austenite, $C'_e$, upon approaching the $M_s$ temperature during cooling even when the intrinsic modulus $C'$ is a constant.

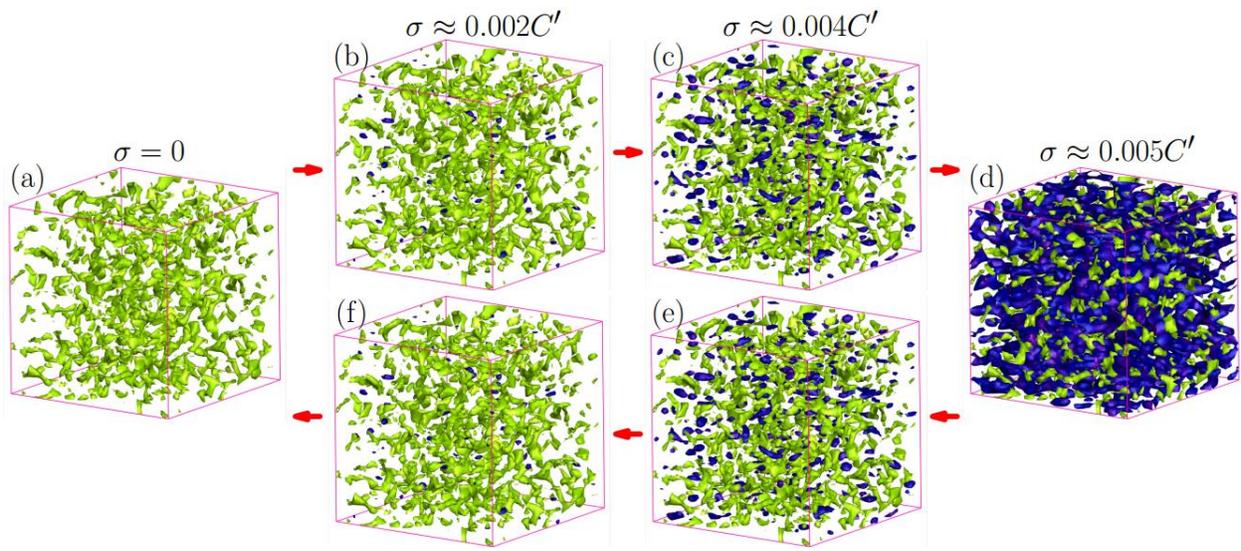

**Figure 9**: The simulated cycle of evolution of the nano-embryonic martensitic structures in the pre-transitional austenite with nano-size precipitates. The evolution is caused by quasi-static external stress that is cyclically applied at temperature 215K. The precipitates were obtained by simulating early stages of decomposition of the pre-transitional austenite. The volume fraction of precipitate phase is about 50%. The assumed dilatation eigenstrain for precipitates is $\varepsilon_p = 2\%$. Red arrows indicate the sequence of the structures determined by cyclically changing applied shear stress.

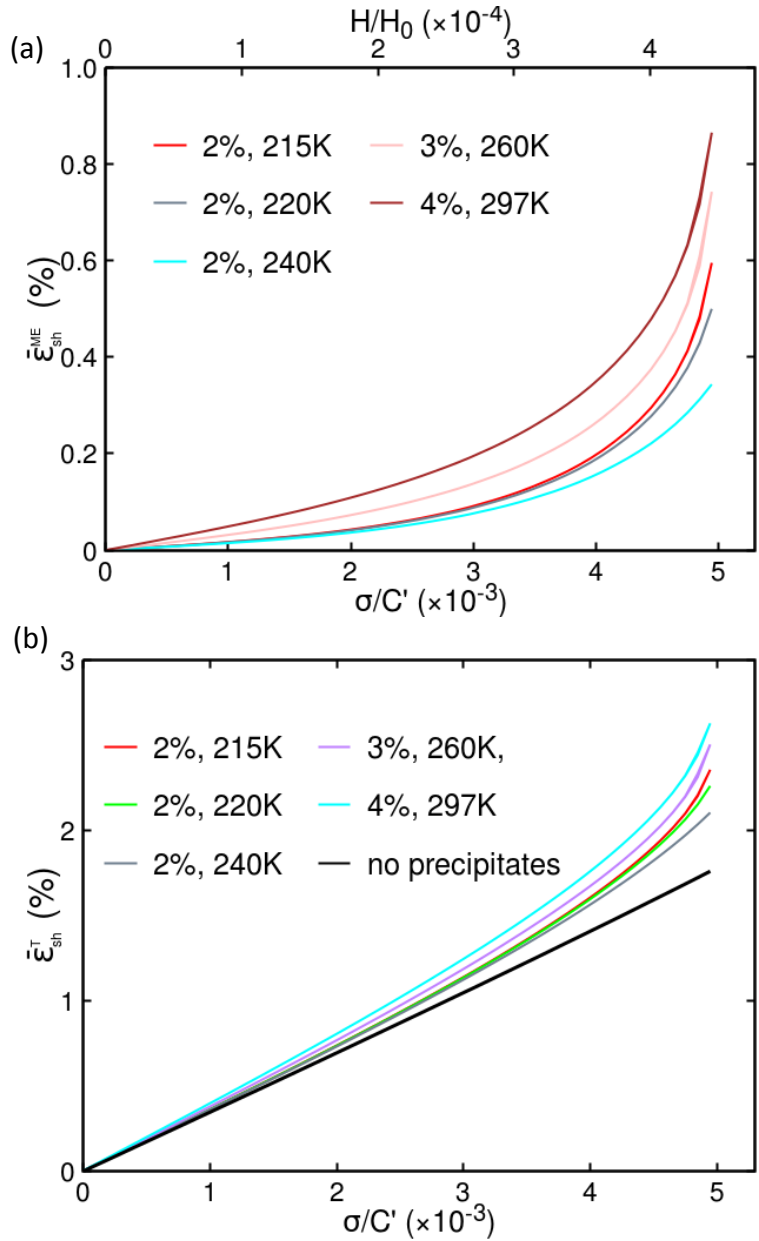

**Figure 10:** Simulated stress-strain dependences for pre-transitional austenites with ~50% precipitates: (a) the shear strain responses provided only by the change of MEs, and (b) the total shear strain responses to a quasi-statically applied shear stress. The precipitates configuration is obtained by simulating decomposition. The used values of dilatational eigenstrains of precipitates, $\varepsilon_p$, and different temperatures are shown in the figure. In (a), the reduced magnitudes of the applied stress and magnetic field are shown on the lower and upper abscissas, respectively. In (b), the black line describes the Hookean strain response (without the contribution from the change of MEs).